\documentstyle[11pt,aaspp4]{article}

\lefthead{Seaquist, Frayer and Bell}
\righthead{${\rm HCO^+}$ in the Starburst Galaxy M82}

\input{psfig}
\input{epsf.sty}
\begin{document}

\title{${\rm \bf HCO^+}$ in the Starburst Galaxy M82}

\author{E. R. Seaquist and D. T. Frayer\altaffilmark{1}}
\affil{Department of Astronomy, University of Toronto, Toronto, Ontario
M5S 3H8}

\and

\author{M. B. Bell}
\affil{National Research Council of Canada, 100 Sussex Drive, Ottawa, Ontario
K1A 0R6}

\altaffiltext{1}{Current address: Department of Astronomy, 105-24 Caltech, Pasadena, CA 91125}

\begin{abstract}

We report observations and an analysis of the distribution of
${\rm HCO^+}$(1-0) and ${\rm HCO^+}$(4-3) emission in the central 1 kpc star forming
region of M82. Comparisons are made with other star formation
indicators such as the mm continuum, the distribution of radio SNR's
and the molecules CO and OH.  In a broad sense, the ${\rm HCO^+}$ is
distributed in a way similar to the CO, although there are noticeable
differences in detail, including an inward displacement of spiral arm
emission relative to CO. A comparison of the position-velocity plots
for CO, ${\rm HCO^+}$(1-0), ${\rm HCO^+}$(4-3), and ionized gas, with orbits expected in
the presence of the nuclear bar suggest an inward transfer of gas
associated with star formation toward the nucleus.

The ${\rm HCO^+}$ (4-3)/(1-0) line ratios are comparatively uniform in the
observed region, and according to an LVG analysis, reflect mean gas
densities in the range ${\rm 10^4 cm^{-3}}$ - ${\rm 10^5 cm^{-3}}$ for kinetic temperatures in
the range 20K - 60K. The comparative uniformity of these conditions
and the low filling factor suggest that each sampled point comprises a
large number of clouds occupying a broad range of density, and
possibly temperature. We briefly examine fractal type models in the
context of the ${\rm HCO^+}$ data as an alternative way to analyze molecular
line emission in M82.

\end{abstract}


\section{Introduction}

The starburst galaxy M82 continues to be the focus of many studies
intended to elucidate the star forming mechanism.  Its proximity (3.25
Mpc) and intense nuclear star formation activity in the inner one kpc
makes it an ideal candidate for this purpose. Much attention has been
focused on the properties of the molecular clouds in this region, and
this topic is also the focus of the present paper. According to the
schematic in Figure 13 of Achtermann \& Lacy (1995), for example, the
molecular gas is concentrated in a ring of radius about 210 pc, and
there is an inner ring of ionized gas with radius of about 80 pc. On
the basis of the position-velocity ({\em p-v}) maps for CO, Shen \& Lo (1995)
have identified two possible spiral arms, one at radius 125 pc, and
the other at 390 pc. Based on their CO interferometric maps, Shen and
Lo found that the molecular gas comprises about 60 molecular clouds
with diameters ranging from 10 to 100 pc, with velocity widths ranging
upward from 7 ${\rm km\,\,s^{-1}}$, similar to clouds in our Galactic center. Based
on interferometric studies of denser gas reflected in the distribution
of HCN, Brouillet \& Schilke (1993) found similar cloud properties, and
also noted their similarity to Giant Molecular Cloud complexes in our
Galaxy.

Here we extend the investigation of the properties of the molecular
gas in M82 by reporting and discussing the distribution of ${\rm HCO^+}$ in
this region. The molecule ${\rm HCO^+}$ has a comparatively high critical
density (approximately ${\rm 10^5 cm^{-3}\, and\, 10^7 cm^{-3}}$ for the (1-0) and (4-3)
transitions respectively). Thus this molecule is preferentially
excited by denser molecular gas, comparable to that responsible for
the excitation of HCN. In addition, however, the ${\rm HCO^+}$ intensities may
be affected by cosmic ray ionization and its excitation may depend on
a number of factors which would not affect HCN (e.g. Richardson et
al. 1988). We have mapped this molecular ion in the (1-0) and (4-3)
transitions, representing the initial phase of a larger study of
tracers of dense gas in M82. This study complements other molecular
line studies as well as our earlier investigation of the ${\rm H}41\alpha$
recombination line, which revealed structures associated with dense
ionized gas (Seaquist et al. 1996). Some of these structures appear to
participate in the outflow from the starburst region. These
observations also revealed velocity displacements between the ionized
and ${\rm HCO^+}$ emitting gas, either a further indication of outflow or
motions by the two species on different orbits. Although previous
papers have reported measurements of both ${\rm HCO^+}$ transitions (e.g. Stark
\& Wolff 1979; Carlstrom 1988; Rieu, Nakai \& Jackson 1989;
Wild et al. 1992), the results reported here are the first to provide
a comparison of ${\rm HCO^+}$ at the (1-0) and (4-3) transitions based on fully
sampled maps.

As part of this study, we also examine the relationship between the mm
radio continuum and the molecular gas inferred from both ${\rm HCO^+}$ and CO
in order to investigate the Schmidt law relating star formation rate
to gas density. A particular benefit of using the mm continuum, which
measures the star formation rate through its association with thermal
emission from ionized gas, is its freedom from obscuration by dust in
M82.

\section{Observations}
     
Observations of the ${\rm HCO^+}$(1-0) line were made with the five element mm
array of the Owens Valley Radio Observatory (OVRO)\footnote{The OVRO Millimeter Array is operated as a radio
astronomy facility by the California Institute of Technology.} during four
observing periods throughout 1993-94, and observations of the  ${\rm HCO^+}$
(4-3) line were made with the James Clerk Maxwell Telescope (JCMT)\footnote{The JCMT is operated by the Joint Astronomy
Centre in Hilo, Hawaii on behalf of the parent organizations PPARC in
the United Kingdom, the National Research Council of Canada, and the
Netherlands Organization for Scientific Research.}
during 1995 December.

\subsection{The ${\rm \bf HCO^+(1-0)}$ observations}

These data were acquired at OVRO simultaneously with data on the
${\rm H}41\alpha$ line, which were reported by Seaquist et al. (1996). The
reader is referred to this paper for details on the observational
procedures. The  ${\rm HCO^+}$(1-0) line at 89.1885 GHz was observed in the
lower sideband, and the ${\rm H}41\alpha$ line at 92.0344 GHz in the upper
sideband. The dates and lower sideband system parameters for these
observations are shown in Table 1. Three different array
configurations were used over the four observing periods, providing a
resultant angular resolution of approximately 3\farcs5. The spectrometer
combined four bands of 32 independent spectral line channels each,
after on-line Hanning smoothing. After deletion of two channels at the
edge of each band, the final number of channels was 112 spread over a
bandwidth of 448 MHz, providing a channel separation of 4 MHz. In
terms of velocity, the total coverage was 1408 ${\rm km\,\,s^{-1}}$ with a channel
separation of 13.46 ${\rm km\,\,s^{-1}}$.  Weather conditions were generally good to
excellent during the observations with the exception of the last
session on 1994 April 28. The rms pointing error was about 4\arcsec \, in each
coordinate except for sunrise and sunset, when the errors were
larger. In all cases the pointing errors were small compared to the
size of the primary beam which is about 88\arcsec.

\placetable{tab1}

\subsection{The ${\rm \bf HCO^+(4-3)}$ observations}

Observations were made of the  ${\rm HCO^+}$(4-3) line at 356.7343 GHz on 1995
December 2 and 3 using receiver B3i on the JCMT. The observations were
conducted with a spectrometer setup providing 1216 channels covering a
bandwidth of 760 MHz and a channel separation of 625 kHz. These
correspond to a total velocity coverage of 639 ${\rm km\,\,s^{-1}}$ and velocity
resolution of 0.527 ${\rm km\,\,s^{-1}}$ respectively. The beam size of the JCMT at
the frequency of the  ${\rm HCO^+}$(4-3) transition is 14\arcsec (FWHM). Measurements
were made at 30 positions in the central region of M82, using a 7\arcsec \,
grid centered on the (B1950) position ${\rm RA = 09^h 51^m 42\fs34}$, ${\rm Dec = 69\arcdeg 55\arcmin 00\arcsec}$ and aligned with the major axis of M82, which was
assumed to be at PA = 75\arcdeg. The spectrometer band was centered at
V(LSR) = 200 ${\rm km\,\,s^{-1}}$. The data were taken by beam switching in azimuth
using a beam throw of 120\arcsec and a switch rate of 7.813 Hz. This beam
throw is sufficient to position the reference beam well off the source
at all map positions. The average air mass was 1.73 for the
observations, and the corresponding mean system temperature was 1197
K, averaged over the central 75\% of the band. The telescope pointing
was monitored every 1 - 2 hours using a strong continuum source as a
calibrator, typically 3C 273 or the nearby source 0923+392. The rms
pointing error was found to be 2\farcs1 in azimuth and 2\farcs6 in elevation,
both substantially smaller than the beam size. The ${\rm T^*_A}$ scale was
converted to main beam brightness temperature ${\rm T_{MB}}$ using a main beam
efficiency of 0.59.

\section{Results and Analysis}

\subsection{The ${\rm \bf HCO^+(1-0)}$ data}

In this section we present an overview of the results, and defer a
detailed discussion of their interpretation and broader implications
to \S4.

The OVRO line data are presented in Figures 1, 2, and 3. Figure 1 is a
series of channel maps, Figure 2 is a position-velocity ({\em p-v}) plot
(shown together with a similar map for CO), with position measured
along the major axis, assumed to be at PA 75\arcdeg, and Figure 3 shows
the integrated brightness map, ie with the line emission summed over all velocity
channels. For comparison, Figure 3 also shows a CO integrated brightness map adapted
from Shen \& Lo (1995), and a 3 mm continuum map, published by Seaquist
et al. (1996). The CO map was convolved to the resolution of the
 ${\rm HCO^+}$ and 3 mm continuum OVRO maps. Figure 4 is a map of the ratio
( ${\rm HCO^+}$/CO) of the integrated brightness maps in Figure 3. This map shows the
distribution of regions of enhancements (and deficiencies) of  ${\rm HCO^+}$
relative to CO emission, and is relevant for comparisons involving
sites of dense molecular gas or enhancements in the abundance of  ${\rm HCO^+}$.
No correction for primary beam attenuation has been applied to the
 ${\rm HCO^+}$ or continuum maps. The attenuation is estimated to be about 10\%
at the easternmost and westernmost extremities of the M82 emitting
region ( $\pm15\arcsec$ from the phase center).

The channel maps in Figure 1 have an rms of 12 mJy, which is close to
the theoretical value for thermal noise estimated from the known
system temperatures and the total integration time. Figures 1 and 2
both show clearly the effect of rotation of the galaxy, and indicate a
systemic velocity of $225\pm10\, {\rm km\,\,s^{-1}}$, in good agreement with other
determinations (e.g.  Weliachew, Fomalont \& Greisen 1984). The
distribution of emission is broadly similar to that seen in CO (Shen \&
Lo 1995) and HCN (Brouillet \& Schilke 1993), clearly indicating the
presence of spiral features. The peak brightness on these maps is
about 200 mJy per beam area, which for the 3\farcs25 x 3\farcs50 gaussian beam
corresponds to a brightness temperature of 2.9 K.  Thus if the
individual clouds contributing this brightness are assumed to be
optically thick with an excitation temperature of 50 K (the dust
temperature), the dilution factor for the peak emission would be about
0.06. These maps therefore suggest that the  ${\rm HCO^+}$ is in the form of
compact clouds heavily diluted by the telescope synthesized beam.

\subsubsection{Comparison of the brightness scale with single dish data}

In order to provide an assessment of missing flux in the
interferometer maps due to missing short spacings, we made a
comparison between our data and  ${\rm HCO^+}$(1-0) measurements made with the
Nobeyama 45m telescope by Rieu, Nakai \& Jackson (1989). The latter
observations were made with an angular resolution of 23\arcsec \, at five
points along the major axis of M82. The central point is at the
nucleus, and the other points are offset at $\pm10\arcsec$ and $\pm25\arcsec$ along
the major axis from the nucleus. For comparison, we convolved the OVRO
maps to an angular resolution of 23\arcsec \, and from the resulting spectral
line cube generated spectra at the positions corresponding to those of
Rieu et al. The comparison between the two sets of line profiles is
shown in Figure 5. Both sets of spectra are shown in units of main
beam brightness temperature (${\rm T_{MB}}$). A main beam efficiency of 0.60 was
used to convert the values of ${\rm T_A^*}$ plotted by Rieu et al. to the ${\rm T_{MB}}$ scale.

Figure 5 generally shows quite good agreement in the nuclear region
and SW side of M82.  On the NE side the line brightness measured by
the interferometer is weaker than that measured by the 45m telescope,
particularly at the position 25NE. The line profiles were integrated
over the velocity range 0-400 ${\rm km\,\,s^{-1}}$, and the resulting OVRO/Nobeyama
line ratios for the five positions (west to east) are $0.75\pm0.05$,
$0.93\pm0.02$, $0.75\pm0.01$, $0.68\pm0.01$, $0.59\pm0.06$, where the
errors refer only to the noise, and do not include calibration
errors. Figure 5 and the lower ratio on the east side suggests that
there is an extended component on this side with velocities between
180 ${\rm km\,\,s^{-1}}$ and 280 ${\rm km\,\,s^{-1}}$ not sampled by the interferometer. This
result is confirmed by comparison with similar (unpublished)
measurements made by us with the 12m telescope of the National Radio
Astronomy Observatory (NRAO) with a much larger FWHM beamsize of
70\arcsec. Once again, the agreement is relatively good on the west side,
but there is even more missing flux on the east side.  These results
place a limit on the precision achievable in the comparison between
the  ${\rm HCO^+}$(4-3) and  ${\rm HCO^+}$(1-0) measurements which are discussed in more
detail later. However, the agreement between the interferometer and
single dish brightness measurements improve as the resolution
increases, which would be expected if the reason is inadequate
sampling of an extended feature.  We conservatively estimate that
(4-3)/(1-0) line ratios using JCMT and OVRO data smoothed to 14\arcsec \,
resolution would be overestimated by an average of about 25\%.

\subsubsection{Comparison with CO}

Overall, there appears to be a disk or torus of emission whose
thickness is roughly comparable to the beam (50 pc). However, there is
considerable structure in the disk, with many features corresponding
to those seen in CO (Shen \& Lo 1995) and HCN (Brouillet \& Schilke
1993), and also there is a broad correspondence between line and
continuum emission, as may be seen by comparing the maps in Figure 3.
In particular, there is good qualitative correspondence between
features in the  ${\rm HCO^+}$ map and features W2, W3, C1, and E2 (designations
by Shen \& Lo) in the CO map, and interpreted by them as the
projections of rings or spiral arms seen tangentially. There are,
however, some variations in the locations of the peak brightness,
which may reflect real variations due to differences in gas
distribution and/or kinematics. For example, there is a significant
displacement between the peaks for the westernmost bright feature W3
as seen in  ${\rm HCO^+}$ and CO, with the former being displaced closer to the
nucleus compared to the latter, placing it in closer correspondence
with the continuum peak of this arm. This effect shows up most clearly
in the comparisons made in Figure 3 among the different
constituents. Such an effect is consistent with that expected if the
star formation were occurring in preferentially denser gas at the
inner edge of the arm. The same effect is not evident on the east
side, where there is good agreement between the positions of
corresponding features. Thus although there is a general
correspondence between features which measure the distribution of gas,
there is some variation with tracer according to critical density and
sensitivity to star forming activity.

Such displacements are confirmed by a comparison between the  ${\rm HCO^+}$ and
CO {\em p-v} plots shown in Figure 2. Note that the aforementioned peaks
have distinct kinematic signatures. There is a strong correspondence
between these two maps of the distribution of velocity along the major
axis of M82.  Shen \& Lo (1995) attribute the existence of the peaks W3
(west of nucleus) and C1 (near the center) in the integrated brightness maps to
``velocity crowding''. However, the term may be somewhat misleading
since these peaks are visible in the {\em p-v} plots as structures spread
out in velocity. Thus they are in fact produced by large column
densities of gas. The feature C1 is associated with gas with a
velocity gradient which is very steep compared with that of the
general rotation curve. This steeply inclined feature is visible both
in  ${\rm HCO^+}$ and CO, and may be an inner ring or spiral arm. This feature
is also closely associated with a ring of ionized gas in the nuclear
region, as may be seen from the diagonal solid line in Figure 2
representing the model for the ionized ring produced by Achtermann \&
Lacy (1995) based on observations of [\ion{Ne}{2}] at 12.8 \micron.  The
location of this ionized ring is also in good agreement with the
location of ionized gas exhibited by [\ion{N}{2}] 6583, reported by G\"otz
et al. (1990), and with observations of the H41$\alpha$ line by Seaquist
et al. (1996). The ionized ring appears displaced toward lower
velocity (or toward the east in position) relative to its counterpart
in the molecular gas (see also \S4.1).

\subsubsection{Comparison with SNR's}

Figure 1 shows that there is a general correspondence between  ${\rm HCO^+}$
emission and the distribution of compact radio SNR's. The latter are
indicated by crosses, based on positions given by Kronberg, Biermann \&
Schwab (1985). There is an apparent avoidance by SNR's of the central
brightest parts of the strongest emission regions suggesting that star
formation occurs at the periphery of these complexes (see for example
the channel maps for V=86 ${\rm km\,\,s^{-1}}$ and 328 ${\rm km\,\,s^{-1}}$). This effect has also
been pointed out by Shen \& Lo (1995) in the case of CO. Although
similar comparisons have been made with star clusters (e.g. Brouillet
\& Schilke 1993), extinction within M82 in the optical and IR makes the
interpretation less clear in this case.

Our data make it possible to test for a correlation between the sites
of SNR's and the existence of dense or ionized molecular gas indicated
by a preferential enhancement of the  ${\rm HCO^+}$/CO ratio shown in Figure
4. Caution must be exercised in making such a comparison because the
 ${\rm HCO^+}$ and CO maps are based on a different sampling of the u-v plane,
but this map should give a qualitative indication of the sites of
highest gas density. An analysis was performed to determine whether
this ratio is systematically higher at the locations of the SNR's than
elsewhere. The analysis showed that, on the whole, the distribution of
the ratios sampled at the sites of SNR's is not significantly
different than for points selected where there are no SNR's.  Thus
there is no evidence that SNR's are preferentially located at the
sites of densest molecular gas measurable by this ratio at 50 pc
resolution.

A notable exception is the compact continuum source 44.01+59.6, one of
the brightest in M82, and which coincides with the position of the
peak  ${\rm HCO^+}$/CO ratio in Figure 4. The source 44.01+59.6 and the
associated peak in  ${\rm HCO^+}$/CO are located approximately 3\arcsec SE of the
2.2\micron \, peak identified by Dietz et al. (1986), and are therefore
spatially distinct from the presumed IR nucleus. Note that the HCN map
of Brouillet and Schilke (1993) also shows a local peak in brightness
within about 1\arcsec \, of the position of the compact continuum source.

The source 44.01+59.6 has been the subject of a recent suggestion that
it is associated with the AGN in M82 (Wills et al. 1997; Seaquist,
Frayer \& Frail 1997). This view is prompted by its unusual radio
spectrum (with a positive spectral index), a strong low frequency
turnover (Wills et al. 1997), and by the possible detection of an
associated jet-like structure (Wills, private communication). The
identification of 44.01+59.6 as the AGN source is supported by
observations by Seaquist et al. (1997) that the satellite OH lines at
1612 MHz and 1720 MHz associated with this source exhibit strong
conjugate emission/absorption similar to that found in the AGN source
of Cen A (van Langevelde et al. 1995) and in the nuclear region of NGC
253 (Frayer, Seaquist \& Frail 1997). No other source in M82 exhibits
this characteristic, even though some other compact sources are
brighter than 44.01+59.6. The inference is that the continuum source
lies behind (or within) a region of gas with a locally very high
column density of dense molecular gas, which is in agreement with our
 ${\rm HCO^+}$ measurement. Note that a similar association between an AGN and
high density molecular gas, but using HCN as a tracer, is found in the
prototype Seyfert galaxy NGC 1068 (Helfer \& Blitz 1995).  It should be
noted that in M82, however, the high ratio of  ${\rm HCO^+}$/CO could be
attributable to either a high gas density in the nuclear region, or to
ionization effects associated with the suspected AGN source. It should
also be noted that if this source is the AGN, it appears displaced both with
respect to the IR nucleus and with repect to the apparent major axis of
the maps in Figures 3 and 4.

\subsubsection{Comparison with OH}

We have also searched for coincidences between high  ${\rm HCO^+}$/CO ratio and
regions of enhanced 1720 MHz satellite maser emission in the nuclear
region of M82 observed by Seaquist et al. (1997).  Features of this
type are found to be associated with 10\% of radio SNR's in our own
Galaxy (Frail et al. 1996), and with the source Sgr A East in the
nuclear region of our Galaxy (Yusef-Zadeh et al. 1996). The 1720 MHz
emission is believed to be preferentially excited in regions of dense
collisionally excited gas in SNR shocks. Three 1720 MHz sources were
found by Seaquist et al. (1997), omitting one south of the plane of
M82, which may be spurious. None of these three coincide with
observable enhancements in the  ${\rm HCO^+}$/CO ratio. However, we call
attention to feature 2 in Seaquist et al. (1997) which is the most
intense 1720 MHz feature, and is located roughly midway between two
discrete continuum sources. One of these, the brightest and most
compact source 41.95+57.5, is about 1\arcsec NW of the maser. It is also
nearly identical in position and velocity with the main line 1667 MHz
maser m2 identified by Weliachew, Fomalont \& Greisen (1974). Figure 6
shows a comparison among the OH and the  ${\rm HCO^+}$(1-0) profiles (measured
at the position of the 1720 MHz feature). The profiles of all three
features are similar suggesting an association between the dense gas
exciting the 1720 MHz line and the  ${\rm HCO^+}$ line.  The density of ${\rm H_2}$
responsible for the OH emission may therefore be comparable to the
critical density for the  ${\rm HCO^+}$ line, or about ${\rm 10^5 cm^{-3}}$.

\subsubsection{Comparison with 3 mm continuum}

We have also made a quantitative comparison between the 3 mm continuum
and the  ${\rm HCO^+}$(1-0) and CO(1-0) integrated line intensities. The mm
continuum is predominantly free-free emission and is thus sensitive to
the emission measure associated with ionized gas, and thus with star
forming regions. Since the 3 mm emission is optically thin, its
brightness is a measure of the star formation rate integrated along
the line of sight. The concentration of molecular gas is also expected
to be higher in star forming regions, and thus the brightness in  ${\rm HCO^+}$
and CO should reflect the distribution of these regions. Thus a
correlation between continuum and molecular gas would be
expected. However, a general correlation is also anticipated simply
because a higher brightness of all such features is associated with
the nuclear disk (e.g. the brightness for all features goes to zero at
the edge of the disk). Therefore we focus on the search for a
{\em nonlinear relationship} between continuum and molecular line emission
which might be expected in association with the Schmidt Law (see
discussion in \S4.2).

Figure 7 shows four plots, two for each molecule, one at the angular
resolution of our  ${\rm HCO^+}$(1-0) data, and the other smoothed to angular
resolution 10\arcsec. The two plots for CO show a relationship whose scatter
decreases with lower resolution, the effect of smoothing over a
stochastic component, or ``noise''. The ``noise'' is not instrumental, but
represents a real decorrelation in the data on scales less than about
100 pc. This is particularly noticeable in the CO map, where the
amplitude of the noise scales clearly with the average brightness
level. The level of the instrumental noise is indicated by the
negative amplitudes at the origin. The case for  ${\rm HCO^+}$ is less clear,
because the S/N ratio is lower.

The plots based on smoothed data appear to exhibit a nonlinear
relationship between line and continuum brightness. We have therefore
used these plots to search for the parameters of a fit of the form log
${\rm \Sigma_R = constant + N\, log \Sigma_G}$, where ${\rm \Sigma_R}$ and ${\rm \Sigma_G}$ are respectively
the continuum and line brightness (with negative values excluded). A
variety of methods outlined by Akritas and Bershady (1996) were used
to fit the data. These methods allow a solution for the parameter N
and its uncertainty when both independent and dependent variables are
subject to error.  The results for  ${\rm HCO^+}$ and CO are ${\rm N = 1.17\pm0.16}$ and ${\rm N = 1.29\pm0.10}$ respectively. Both plots exhibit ${\rm N > 1}$, signifying
a higher than linear dependence, but only the result for CO appears
significant. We note that a similar power law relationship exists for
${\rm H\alpha}$ surface brightness vs. total gas surface density for other
galaxies, where it is found that ${\rm N \sim 1.3 - 1.4}$ (e.g. Kennicutt 1990,
1998). We return to this point in \S4.2.

\subsection{The ${\rm \bf HCO^+(4-3)}$ data}

The  ${\rm HCO^+}$ (4-3) data obtained with the JCMT are shown in two
forms. Figure 8 shows a grid of all the line profiles together with
the OVRO profiles for  ${\rm HCO^+}$(1-0) at the corresponding positions.  The
latter profiles were generated from the spectral line cube of the OVRO
data after first convolving the latter to the angular resolution of
the  ${\rm HCO^+}$(4-3) data (14\arcsec)  and converting the intensities to main beam
brightness temperatures. Figure 9 shows the {\em p-v} plots for both data
sets along the major axis. Comparisons between these two transitions
must be made with caution, since the OVRO measurements under sample the
short spacings. The analysis in  \S3.1.1 indicates that the OVRO
intensities may be low by an average of 25\%, more so on the east side.
Nevertheless, the comparison provides information on the upper limits
to the (4-3)/(1-0) ratio, and can provide a qualitative picture of any
variations in the line ratio on small spatial scales.

Figures 8 and 9 indicate that there is a broad similarity between the
distributions of  ${\rm HCO^+}$(1-0) and  ${\rm HCO^+}$(4-3) emission implying that the
mean excitation conditions are relatively uniform throughout the
nucleus of M82.  We examine this further by looking at the spatial
variation of the line ratios, always bearing in mind the cautionary
note in the previous paragraph.  Integrated line ratios were computed
for each position in Figure 8 by determining the numerical value of
the scaling factor for the (1-0) profile which produces the best match
to the (4-3) line in the least squares sense. These values are noted
in Figure 8, together with their errors. We have also computed the
mean ratio weighted by the inverse square of the errors, which yields
0.39 with a dispersion (standard deviation) of 0.17. The dispersion is
about twice the mean standard error (0.09) of the individual measured
ratios. Much of the observed dispersion is attributable to a
systematic increase in the ratio with position along the major axis,
increasing from west to east.  However, this trend may be attributable
to under-sampling of the u-v plane suspected from the results of
Figure 5, since there is evidence there that the line flux is
underestimated on the eastern side of M82. Therefore this trend is
probably not real. Note that there is some evidence for a similar
trend in the  ${\rm HCO^+}$(1-0)/CO(1-0) ratio, possibly resulting from a
similar under-sampling problem in the CO(1-0) map.

Finally, a careful examination of the {\em p-v} plots in Figure 9 reveals a
small inward displacement (by a few arcsec) of the peaks representing
outer spiral arm emission (features E2 and W3) in  ${\rm HCO^+}$(4-3) relative
to that in  ${\rm HCO^+}$(1-0). This may be related to a similar effect noted in
\S3.1.2 in connection with  ${\rm HCO^+}$ and CO.  This point is discussed
further in \S4.1.

\section{Discussion}

\subsection{The kinematics of the molecular and ionized gas}

Figure 2 shows the {\em p-v} plots along the major axis for  ${\rm HCO^+}$(1-0) and
CO(1-0), indicating the rotation curves for the molecular gas.  Also
shown superimposed is a line marking the corresponding diagram for the
ionized ring identified from [\ion{Ne}{2}] observations by Achtermann \& Lacy
(1995). There is clearly a strong qualitative similarity between the
distributions of CO and  ${\rm HCO^+}$. This ridge, which is exceptionally
linear and which possesses a large velocity gradient, is responsible
for features C1 and C2 in the central region of the CO integrated brightness map of
Shen \& Lo (1995). The ionized gas feature is more closely aligned with
this ridge than with the general distribution of molecular gas, but
appears blue-shifted with respect to this ridge, by up to 30 ${\rm km\,\,s^{-1}}$. This displacement is confirmed by a similar displacement between
${\rm H}41\alpha$ and  ${\rm HCO^+}$(1-0) noted by Seaquist et al. (1996). The inference
is that the steep ridge is associated with molecular and ionized gas
and star formation. This interpretation is supported by the observed
distribution of 1720 MHz OH line masers associated with radio SNR's in
M82, which also follow this ridge and the zone where  ${\rm HCO^+}$ is most
prominent. The appearance of a comparable structure on the east side
is not evident. The indication is therefore that star formation is
especially active on the west side of M82.

Based on their computations of orbits in the nuclear region,
Achtermann \& Lacy (1995) note the similarity between the {\em p-v} plot of
the ionized ring and the x2 orbits of gas responding to the
gravitational field of the bar, which appear significantly steeper
than the larger x1 orbits defining the solid-body portion of the
rotation curve. The closely associated steep and linear part of the
rotation curve seen in molecular emission probably has a similar
origin. According to this interpretation, the velocity displacements
between ionized and molecular gas would require that the ionized gas
be located in the outermost x2 orbits, perhaps associated with gas
transferred from the inner x1 (intersecting) orbits, signifying
intense star formation caused by gas in the colliding
orbits. Alternatively, G\"otz et al. (1990) attribute the kinematics
seen in [\ion{N}{2}] in part to outflow from the nuclear disk. In the latter
picture, the observed negative displacement in velocity might be
attributable to shocked ionized gas entrained in the outflow (see also
Seaquist et al. 1996).  Note however that their model assumes that the
northern edge of the M82 disk is the near side, whereas Wills
et. al. (1997) show new and convincing evidence that the southern edge
is the near side, based on a study of the radio SNR's. Thus this model
may require some revision.

As noted in \S3.1.2 and \S3.2, the {\em p-v} plots in Figures 2 and 9 reveal
a slight inward displacement (toward the nucleus) of the spiral arm
peaks associated with denser and more highly excited gas relative to
those associated with lower density (features E2 and W3 in the maps of
Shen \& Lo (1995)). The magnitude of this displacement is about 3\arcsec (45
pc). The effect is to produce a slightly steeper velocity gradient
when measured by these features. This effect might indicate that, like
the ionized gas, the denser molecular gas also follows orbits in
transition between the inner x1 and the outer x2 orbits near the inner
Linblad resonance (see Figure 16 of Achtermann \& Lacy 1995). Such gas
would have undergone collisions due to the intersecting x1 orbits
leading to star formation. As noted earlier there is also a
significant displacement inward of  ${\rm HCO^+}$ in the integrated brightness and 3 mm
continuum maps for the feature W3, which is probably related to the
displacements seen in the {\em p-v} plots.

The  ${\rm HCO^+}$ observations thus provide evidence of gas falling inward
toward the nucleus where star formation and hence ionization is more
pronounced. Note that  ${\rm HCO^+}$ may also be excited by other effects
(e.g. ion drift) or enhanced in abundance by ionization effects, each
of which could also be enhanced in the region of transfer between the
x1 and x2 orbits because of shocks created by cloud collisions.

\subsection{The Schmidt Law}

We noted in \S3.1.5 the existence of a non-linear relation between the
mm continuum and molecular line brightness, especially for CO.  The
former quantity is associated with the star formation rate and the
latter with the molecular gas density, both integrated over the line
of sight.  Thus the relationship represents an alternative way of
investigating the Schmidt Law, ie the relationship between the local
star formation rate and the associated gas density.  In its original
form the Schmidt Law relates these quantities by a simple power law ${\rm R
= a(\rho_g)^n}$ (Schmidt 1959), where a is a constant and n = 2. In external
galaxies, the observables are projected surface densities, leading to
the alternate form ${\rm \Sigma_R = A(\Sigma_G)^N}$ . The quantity ${\rm \Sigma_R}$ is measured by the H$\alpha$
surface brightness and ${\rm \Sigma_G}$ is obtained from CO and \ion{H}{1} observations,
which yield the gas column density.

The results based on observations of galaxies are reviewed by
Kennicutt (1990). In the case of studies of spatially resolved
observations of (near face-on) galaxies, ${\rm N \sim 1.3}$ is found for gas
densities exceeding the threshold value ${\rm \Sigma_G \sim 3 M_{\sun} pc^{-2}}$, with the star
formation being totally suppressed at column densities significantly
below this value. New results by Kennicutt (1998) show that for
globally averaged surface brightnesses applied to large samples of
disk galaxies, ${\rm N \sim 1.4}$. In M82 we find  ${\rm N \sim 1.3}$ using the measured CO
brightness as a measure of total gas column density and the mm
continuum surface brightness as a measure of the star formation rate
(instead of H$\alpha$). Neither of the two variables measured are subject
to significant interstellar extinction by dust through the disk. Note
however that M82 is nearly edge-on instead of nearly face-on so that
it is not possible to directly infer the gas column density
perpendicular to the disk. The column density measured perpendicular
to the disk may however be estimated from the observed column density
using a simple uniform disk model. The results indicate that the gas
column density perpendicular to the disk significantly exceeds the
threshold value shown above.

We have also investigated whether N = 1.3 is a reasonable value,
assuming that the Schmidt Law with n = 2 is valid, and that the
observed surface densities are represented by integrals over the
distributions of star formation rate and gas density. The effect of
integrating along the line of sight is to sum the volume density of
the star formation rate and gas over a statistical ensemble of
variations in path length. The quantity N can be computed analytically
if it is assumed that the path length is statistically correlated with
the gas density (e.g. shorter path lengths are associated with higher
density). If for example, we take n = 2, and represent the
distribution of mean path length ${\rm L}$ as a gaussian distribution of
density (ie ${\rm L\, \, \propto\, e^{-const(\rho^2)}}$) then we find N = 1.5, comparable
to the value observed. Note that the value of N is not determined by
whether the galaxy is nearly face-on or nearly edge-on, but instead by
the form of the statistical relationship between mean path and density
distribution in the regime above the critical surface mass density.

\subsection{Modeling the line ratios}

The line ratios shown in Figure 8 in principle yield information on
the excitation conditions in the molecular gas.  Our  ${\rm HCO^+}$ data alone
are not sufficient to determine unique values of the excitation
conditions, In addition, as noted in \S4.2, the line ratios are
probably overestimated by about 25\% because of incomplete sampling of
the u-v plane. We therefore restrict our discussion instead to issues
which do not depend critically upon the precise absolute numerical
ratios, and to illustrative physical conditions. Figure 8 shows that
the ratios vary comparatively little over the nuclear region, and are
all (within the errors) equivalent to the weighted mean value of about
0.4.  Since the region observed would be expected to encompass a wide
range of physical conditions, each sampling point must cover a
similarly wide range in physical conditions, represented by an
ensemble of a large number of molecular clouds, in which the average
cloud is subthermally excited.

\subsubsection{Conditions in the average  ${\rm HCO^+}$ emitting cloud}

To model illustrative physical conditions (averaged over the beam)
satisfying the line ratios, we have assumed that the excitation is
determined by the molecular abundance, gas density and temperature
only, and have used a standard LVG approach to compare the expected
line ratios with the ones observed.  The collision cross sections used
for  ${\rm HCO^+}$ are from Monteiro (private communication), and include levels
up to J = 11, with a correction for helium. We therefore ignore a
variety of other effects which could influence the line ratios for
 ${\rm HCO^+}$, discussed by Richardson et al. (1988). These other effects
include collisions by electrons, ion drift (ie velocity differences
between ionized and neutral species in the presence of a magnetic
field), and radiative excitation by the mm continuum radiation field
in M82. The LVG computations require knowledge of the parameter ${\rm X
(dv/dr)^{-1}}$, where X is the abundance of  ${\rm HCO^+}$ relative to ${\rm H_2}$, and dv/dr
is the velocity gradient in the clouds. A number of analyses of  ${\rm HCO^+}$
have been performed in molecular clouds in our own
Galaxy. (e.g. Richardson et al. 1986; Richardson et al.  1988; Heaton
et al. 1993; Irvine et al. 1987). Derived values for X fall in the
range ${\rm 1.0x10^{-9} < X < 1.0x10^{-8}}$. Values of dv/dr (in ${\rm km\,\,s^{-1} pc^{-1}}$) are
generally near unity. For our calculations we employed ${\rm X (dv/dr)^{-1} =
1.0x10^{-9}}$. By computing line ratios for a series of densities and
kinetic temperatures in the range ${\rm 10^2 cm^{-3} - 10^8 cm^{-3}}$, and ${\rm T_{kin} = 20K}$,
40K, and 60K, we find that the $\pm1\sigma$ range in line ratios spanned
by the observations $(0.39\pm0.17)$ is produced by densities in the
range ${\rm 1.8x10^4 cm^{-3} - 1.6x10^5 cm^{-3}}$. This range of densities is
comparable to those found by Seaquist et al. (1996) for the most dense
ionized regions using the ${\rm H}41\alpha$ line as a diagnostic.

In principle, the model brightness temperatures can be combined with
the observed brightness temperatures to infer the geometric dilution
factor F, which is determined by the product of the area covering
factor and the velocity dilution factor. This would require the
development of models involving more transitions to uniquely determine
the densities and temperatures. However, we can again look at a set of
illustrative conditions. For example, the peak (1-0) line brightness
temperature in Figure 8 is about 0.6K, just east of the IR nucleus,
where the line ratio is about 0.33. If the kinetic temperature is 40K,
the LVG models yield a corresponding density of about ${\rm 3x10^4 cm^{-3}}$ and a
model brightness temperature of 15K, leading to a dilution factor of
0.04. Thus the slice through the image cube at this position must
reflect a large number of clouds filling about 4\% of the cube volume.

\subsubsection{${\rm HCO^+}$ and models with fractal geometry}

The range of densities derived above ${\rm (10^4 cm^{-3} - 10^5 cm^{-3})}$ could be
thought of as representing a set of mean conditions reflecting the
 ${\rm HCO^+}$ emitting regions. The gas is likely to be very nonuniform in
density, and a different approach using models with multiple
components will ultimately be necessary for the interpretation of
multi-line data. The substantial uniformity of the line ratios implies
that the beam does not resolve these components, and that each
sampling point is measuring average conditions over an ensemble of
compact clouds, whose statistical properties are quite uniform from
point to point. Quiescent molecular clouds in our Galaxy appear to
exhibit fractal structure incorporating a power-law for the number
density - mass relationship. As noted by Elmegreen \& Falgarone (1996),
an idealized (fractal) sample including all clouds and cloud clusters
in the hierarchy should give a cloud mass spectrum with a power law
index $\alpha = -2.0$.  Power law relationships are also observed for the
dependence of the average density and velocity dispersion of the
clumps on their size (e.g. Zinnecker 1990). The range in cloud size
covered by these relationships is about ${\rm 10^{-2} pc\,\, to\,\, 10^{2} pc}$, ie four
orders of magnitude. These power law relationships are believed to be
governed either by turbulence or self-gravitation. This structure may
be causally related to the form of the initial mass function for stars
(e.g. Elmegren \& Falgarone 1996; Elmegren 1997).

Although a detailed discussion of such models is premature and beyond
the scope of this paper, we report that in a preliminary investigation
of such a model, we find that the  ${\rm HCO^+}$ line ratios are compatible with
$\alpha = -2.0$, when the other power law relationships are constrained to
those characterizing quiescent interstellar clouds in our own
Galaxy. Such models will be interesting to pursue when sufficient line
data become available. We will discuss such models and their relevance
in a forthcoming paper.

\section{Summary and Conclusions}

The analysis of  ${\rm HCO^+}$ observations for M82 shows that the  ${\rm HCO^+}$(1-0)
emission in M82 is distributed throughout the inner 1 kpc in a way
broadly similar to that of CO, and reflects the distribution of star
forming gas. There are some differences with respect to CO which
reflect its excitation by higher gas density, and possibly other
effects such as ionization by cosmic rays or UV emission in star
forming regions. The  ${\rm HCO^+}$(1-0) tends to concentrate more toward the
inner edges of the spiral arms, similar to the ionized gas.  This star
forming gas may represent material in the x2 orbits which are near the
inner Lindblad resonance and which are associated with the
gravitational influence of the bar in M82. This gas may be the product
of transfer from the x1 orbits which have larger radii.

The distribution of  ${\rm HCO^+}$(4-3) is similar to that of  ${\rm HCO^+}$(1-0), except
for a slight inward displacement of the material in the spiral arms,
as revealed by the {\em p-v} plots. This inward displacement, together with
a similar displacement of  ${\rm HCO^+}$(1-0) relative to CO in the spiral arms
may be evidence for gas in transition between the x1 and x2 orbits,
signifying an inward transfer of gas. The  ${\rm HCO^+}$ (4-3)/(1-0) intensity
ratio is relatively constant over the extent of the inner kpc region,
indicating mean ${\rm H_2}$ densities in the range ${\rm 10^4 cm^{-3} - 10^5 cm^{-3}}$ for
kinetic temperatures in the range 20K- 60K. This is similar to the
maximum densities derived for ionized gas from recombination line
studies.

The low filling factor and the relatively constant  ${\rm HCO^+}$ (4-3)/(1-0)
line ratio suggest that each point in the map contains a large
ensemble of clouds with relatively uniform statistical properties. A
preliminary application of fractal-type models based on studies of
quiescent molecular clouds in our own Galaxy shows the observed line
ratios are consistent with the mass spectrum of quiescent molecular
clouds, provided the physical gas density scales with the mean cloud
density.

There is no apparent enhancement in  ${\rm HCO^+}$ relative to CO at the sites
of the compact radio emitting SNR's in M82, except for the source
44.01+59.6, where the  ${\rm HCO^+}$/CO brightness ratio is exceptionally
high. This supports other evidence from continuum and OH satellite
line studies that this source is associated with an AGN in M82.

Finally, an investigation of the Schmidt Law for star formation, using
the 3 mm radio continuum as a measure of star formation rate, and
primarily CO brightness as a measure of gas column density, shows that
N = 1.3, commonly found for nearly face-on spirals, is also consistent
with the behavior of the nuclear region of M82. The inclination of the
galaxy is likely to be unimportant in determining this parameter
provided the measured quantities are unaffected by absorption by dust
in M82.

\acknowledgments
{
ERS acknowledges the support of a grant from the Natural Sciences and
Engineering Research Council of Canada. We thank the staffs of the
Owens Valley Radio Observatory and the James Clerk Maxwell Telescope
for their assistance, and we thank Ms. Sandra Scott for help with the
preparation of this manuscript.
}

\clearpage

\begin{figure}
\epsfbox{fig1a.epsi}
\end{figure}
\pagebreak
\begin{figure}
\epsfbox{fig1b.epsi}
\end{figure}
\newpage
\begin{figure}
\epsfbox{fig1c.epsi}
\end{figure}
\newpage
\begin{figure}
\epsfbox{fig1d.epsi}
\figcaption{fig1a-d}{Channel maps of  ${\rm HCO^+}$(1-0) emission in
M82. Crosses refer to the locations of the radio SNR's (Kronberg et al.
1985), with size roughly proportional to the flux,  and the diamond
marks the position of the IR nucleus according to Dietz et al. (1986).
The contour levels in flux density per beam (main beam temperature) are
\mbox{35 mJy (0.473 K) x (-1,1,3,5,7).} The last image is the continuum at 92
GHz, and shows the beamsize for all maps at the lower left corner. The
contours for the continuum map are \mbox{3.0 mJy x
(-1,1,2,3,4,6,8,10,12,16,20). } }
\end{figure}
\newpage

\begin{figure}
\epsfbox{fig2.epsi}
\figcaption{}{Position-velocity plots of   ${\rm HCO^+}$(1-0)
(top) and CO(1-0) (bottom) emission in M82. The latter plot is adapted
from  Shen \& Lo (1995), convolved to the angular resolution of the
${\rm HCO^+}$ data (3\farcs50 x 3\farcs25). In each case the horizontal
axis corresponds to angular displacement (east is positive) along the
major axis at PA = $75\arcdeg$ passing through the IR nucleus defined by Dietz
et al.  (1986). The reference coordinates correspond to the location of
the IR nucleus and VLSR = 200 ${\rm km\,\,s^{-1}}$. The contours are at
uniform spacing with interval 0.406K for  ${\rm HCO^+}$ and 4.86K for
CO.  The diagonal line marks the location of the model for the ionized
ring by Achtermann \& Lacy (1995).}
\end{figure}

\begin{figure}
\epsfysize=8.0in \epsfbox{fig3.epsi}
\figcaption{}{Three maps, each at angular resolution 3\farcs50 x
3\farcs25 showing the brightness integrated over velocity for ${\rm HCO^+}$(1-0) with
uniform contours at intervals of 40.6 K-km ${\rm s^{-1}}$ (top), the
integrated brightness for CO(1-0) with uniform contours at intervals of
324 K-km ${\rm s^{-1}}$, adapted from Shen \& Lo (1995) (middle), and
3mm continuum, with contours as in Figure 1 (bottom). }
\end{figure}


\begin{figure}
\epsfbox{fig4.epsi}
\figcaption{}{A map of the ratio of the (ie  ${\rm HCO^+}$/CO)
using the maps in Figure 3, showing also the distribution of radio
SNR's (crosses) and the IR nucleus, as in Figure 1. Regions where the
CO emission  is fainter than 3 are blanked. The contours are uniform
with an interval of 0.05. The grey scale axis shows the milliratio.}
\end{figure}


\begin{figure}
\epsfysize=8.0in  \epsfbox{fig5.epsi}
\figcaption{}{A comparison between line profiles of  ${\rm
HCO^+}$(1-0) measured by Rieu et al. (1989) using a single aperture
radio telescope (thick lines) and corresponding profiles from our own
interferometer measurements (thin lines) at five different positions in
M82. The interferometer data cube was convolved to the same angular
resolution (23\arcsec) as the single dish measurements for the purpose
of generating the profiles based on OVRO data. }
\end{figure}


\begin{figure}
\epsfysize=8.0in  \epsfbox{fig6.epsi}
\figcaption{}{Comparison of spectral profiles for   ${\rm
HCO^+}$(1-0) from this paper (top), OH (1720 MHz) from Seaquist et al.
1997, (middle), and OH (1667 MHz) from Weliachew et al. (1984) for the
OH maser region near the radio SNR 41.95+57.5 (bottom). The  ${\rm
HCO^+}$ measurements have an angular resolution of 3\farcs 5 whereas
the the OH measurements have an angular resolution of approximately
1\arcsec.}
\end{figure}


\begin{figure}
\epsfysize=8.0in \epsfbox{fig7.epsi}
\figcaption{}{Plots of  3mm continuum brightness (vertical
scale) vs. integrated  ${\rm HCO^+}$(1-0) and CO(1-0) emission
(horizontal scale).  Plots (a) and (c) were made using maps with the
original angular resolution of the  ${\rm HCO^+}$(1-0)  measurements,
and (b) and (d) were made using maps convolved to an angular resolution
of 10\arcsec. Curves shown for the latter two plots are of the form 
${\rm \Sigma_R = A(\Sigma_G)^N}$, where N = 1.17 for (b) and N = 1.29 for (d).}
\end{figure}


\begin{figure}
\epsfysize=8.0in \epsfbox{fig8.epsi}
\figcaption{}{Map showing   ${\rm HCO^+}$(1-0) and  ${\rm
HCO^+}$(4-3) profiles for M82, and the ratios (4-3)/(1-0) of the
integrated intensities, with their errors (see text for details). The
data cube for the (1-0) transition was convolved to the same angular
resolution as that for the (4-3) transition (14\arcsec) before
constructing the profiles. The coordinates for each panel refer to the
origin of the plot.}
\end{figure}


\begin{figure}
\epsfbox{fig9.epsi}
\figcaption{}{Position-velocity plots showing the distribution
of  ${\rm HCO^+}$(4-3) emission (top) shown in contour form compared
with a similar plot of   ${\rm HCO^+}$(1-0) emission (bottom). The gray
scale image associated with the  ${\rm HCO^+}$(4-3) plot represents
the  ${\rm HCO^+}$(1-0) emission in the lower panel for comparison.
The  ${\rm HCO^+}$(1-0) emission was convolved to the angular
resolution of the  ${\rm HCO^+}$(4-3) data (14\arcsec). Contours for
both plots are at uniform intervals of 0.04K in ${\rm T_{MB}}$.}
\end{figure}

\begin{table} 

\caption{PARAMETERS FOR THE OVRO OBSERVATIONS\label{tab1}}

\begin{tabular}{ccccc}
&     Date          &Baselines      &${\rm <T_{sys}>}$         &Integration \\
&      &                  (m)                &(K)          &(minutes) \\
1993 &Nov 15                &20-67          &484               &525 \\

1993 &Dec 23                &40-206         &449               &380 \\

1994 &Jan 29                &15-119         &508               &645 \\

1994 &Apr 28                &20-67          &833               &425 \\

\end{tabular}
\tablecomments{${\rm <T_{sys}>}$ means system temperature averaged over time and sideband.}

\end{table}



\begin{references}
\reference{}Achtermann, J. M., \& Lacy, J. H. 1995, \apj, 439, 163
\reference{}Akritas, M. G., \& Bershady, M. A. 1996, \apj, 470, 706
\reference{}Brouillet, N., \& Schilke, P. 1993, \aap, 277, 381
\reference{}Carlstrom, J. E. 1988, in Galactic and Extragalactic Star Formation, eds. R.E. Pudritz, \& M. Fich (Dordrecht:Reidel), 571
\reference{}Dietz, R. D., Smith, J., Hackwell, R. D., Gehrz, R. D., \& Grasdalen, G. L. 1986, \aj, 91, 758
\reference{}Elmegreen, B. G. 1997, \apj, 486, 944
\reference{}Elmegreen, B. G., \& Falgarone, E. 1996, \apj, 471, 816
\reference{}Frail, D. A., Goss, W. M., Reynoso, W. M., Giacani, E. B., Green, A. J., \& Otrupeck, R. 1996, \aj, 111, 1651
\reference{}Frayer, D. T., Seaquist, E. R., \& Frail, D. A. 1998, \aj, 115, 559
\reference{}Heaton, B. D., Little, L. T., Yamashita, T., Davies, S. R., Cunningham, C. T., \& Monteiro, T. S. 1993, \aap, 278, 238
\reference{}Helfer, T. T. \& Blitz, L. 1995, \apj, 450, 90
\reference{}Irvine, W. M., Goldsmith, P. F., \& Hjalmarson, \AA. 1987, in Interstellar Processes, eds. D. J. Hollenbach, \& H. A. Thronson (Dordrecht:Reidel), 585
\reference{}G\"{o}tz, M., McKeith, C. D., Downes, D., \& Greve, A. 1990, \aap, 240, 52
\reference{}Kennicutt, R. C., Jr. 1990,  in The Interstellar Medium in Galaxies, eds. H. A. Thronson, Jr., \& J. M. Shull (Dordrecht:Kluwer), 405
\reference{}Kennicutt, R. C., Jr. 1998, \apj, in press
\reference{}Kronberg, P. P., Biermann, P., \& Schwab, F. R. 1985, \apj, 291, 693
\reference{}Richardson, K. J., White, G. J., Monteiro, T. S., \& Hayashi, S. S. 1988, \aap, 198, 237
\reference{}Richardson, K. J., White, G. J., Phillips, J. P., \& Avery, L. W. 1986, \mnras, 219, 167
\reference{}Rieu, N. Q., Nakai, N., \& Jackson, J. M. 1989, \aap, 220, 57
\reference{}Schmidt, M. 1959, \apj, 129, 243
\reference{}Seaquist, E. R., Carlstrom, J. E., Bryant, P. M., \& Bell, M. B. 1996, \apj, 465, 691
\reference{}Seaquist, E. R., Frayer, D. T., \& Frail, D. A. 1997, \apj, 487, L131
\reference{}Shen, J. \& Lo, K. Y. 1995, \apj, 445, L99
\reference{}Stark, A. A., \& Wolff, R. S. 1979, \apj, 229, 118
\reference{}van Langevelde, H. J., van Dishoeck, E. W., Sevenster, M., \& Israel, F. P. 1995, \apj, 448, L119
\reference{}Weliachew, L., Fomalont, E. B., \& Griesen, E. W. 1984, \aap, 137, 335
\reference{}Wild, W., Harris, A. I., Eckart, A., Genzel, R., Graf, U. U., Jackson, J. M., Russell, A. P. G., \& Stutski, J. 1992, \aap, 265, 447
\reference{}Wills, K. A., Pedlar, A., Muxlow, T. W. B., \& Wilkinson, P. N. 1997, \mnras, 291, 517
\reference{}Yusef-Zadeh, E., Roberts, D. A., Goss, W. M., Frail, D. A., \& Green, A. J. 1996, \apj, 466, L25
\reference{}Zinnecker, H. 1990, in Physical Processes in Fragmentation and Star Formation, eds. R. Capuzzo-Dolcetta et al. (Dordrecht:Kluwer), 201 

\end{references}
\end{document}